\documentclass[preprintnumbers,amsmath,amssymb,floatfix,10pt,prd,twocolumn,superscriptaddress,nofootinbib]{revtex4}
\usepackage[colorlinks=true, pdfstartview=FitV, linkcolor=magenta, citecolor=red, urlcolor=blue]{hyperref}
\usepackage{epsfig,color}
\usepackage{xcolor}
\usepackage{epstopdf}
\usepackage{amssymb}
\usepackage{verbatim}
\usepackage{multirow}
\usepackage{latexsym}
\usepackage{epsfig}
\usepackage{epstopdf}
\usepackage{graphicx}
\usepackage{amssymb}
\usepackage{amsmath}
\usepackage{dcolumn}
\usepackage{bm}
\usepackage{color}
\usepackage{comment}

\begin{document}
\title{Study on physical properties and maximum mass limit of Finch-Skea anisotropic model under Karmarkar condition in $f(Q)$-gravity}

\author{G. Mustafa}
\email{gmustafa3828@gmail.com} 
\affiliation{Department of Physics, Zhejiang Normal University, Jinhua 321004, Peoples Republic of China,}
\affiliation{New Uzbekistan University, Mustaqillik ave. 54, 100007 Tashkent, Uzbekistan,}

\author{Allah Ditta}%
\email{mradshahid01@gmail.com}
\affiliation{Department of Mathematics, Shanghai University  and Newtouch Center for Mathematics of Shanghai University,  Shanghai, 200444, People's Republic of China}

\author{Saadia Mumtaz}
\email{saadia.icet@pu.edu.pk}
\affiliation{Institute of Chemical
Engineering and Technology, University of the Punjab, Quaid-e-Azam Campus, Lahore-54590, Pakistan}

\author{S.K. Maurya}
\email{sunil@unizwa.edu.om}\affiliation{Department of Mathematical and Physical Sciences, College of Arts and Sciences, University of Nizwa, Nizwa 616, Sultanate of Oman}

\author{De\u{g}er Sofuo\u{g}lu}
\email{degers@istanbul.edu.tr}          
\affiliation{Department of Physics, Istanbul University, 34134, Vezneciler, Fatih, Istanbul, Turkey}

\begin{abstract}
The primary objective of this work is to study the dynamical characteristics of an anisotropic compact star model with spherical symmetry. This investigation is conducted in the framework of $f(Q)$ modified gravity. To simplify the calculations, we employ the Karmarkar condition and derive a differential equation that establishes a relationship between two crucial components of the spacetime namely $e^\nu$ and $e^\lambda$. Additionally, we incorporate the well-known Finch-Skea structure as the component representing $g_{rr}$ and subsequently find the resulting form of the component $g_{tt}$ from the relation of metric functions to formulate the precise solutions for the stellar structure.  To assess the behavior of the anisotropic fluid and stability of the compact star, we use the observed values of mass and radius for the compact star model $PSR J0437-4715$. The graphical analysis depicts that the stellar structure possesses physical viability and exhibits intriguing properties. Furthermore, we predicted the mass-radius relation along with the maximum mass limit of several objects for different parameter values by assuming two different surface densities. It is discovered that the compactness rises when density increases.\\

\end{abstract}

\date{\today}

\maketitle

\section{introduction}\label{sec1}

It has become evident through various cosmological phenomena \cite{Riess98, Perlmutter99, Ade16, Aghanim16} that the modifications to general relativity (GR) are imperative at a geometric scale. General relativity, which is based on Riemannian geometry, describes the Levi-Civita connection and considers the Ricci curvature $R$ as the fundamental property of spacetime. This formulation excludes torsion and non-metricity from the geometry. An alternative approach to describe GR, alongside Riemannian geometry, is teleparallel gravity \cite{Aldrovandi14}. In this theory, the gravitational force arises from torsion $\mathcal{T}$ instead of curvature $R$ in the Riemannian geometry. This concept was formally introduced by Einstein who previously employed teleparallel geometry in his endeavors to develop a unified field theory \cite{Einstein28}. Another approach, known as non-metricity or symmetric teleparallel gravity, has been proposed by Nester \cite{Nester99}. In this approach, the non-metricity $Q$, which is independent of curvature and torsion, serves as the mediator of gravitational interactions. Unlike GR, where gravity is mediated by the affine connection rather than the physical manifold, non-metricity plays a crucial role in this alternative approach.

The Einstein pseudo-tensor, which becomes a true tensor in the geometric representation, corresponds to the non-metricity $Q$. Jimenez et al. \cite{Jimenez18} extended symmetric teleparallel gravity to coincide with GR, resulting in $f(Q)$ gravity. In light of the increased emphasis on extended gravity theories within contemporary cosmology, researchers are investigating alternative geometries with the aim of attaining a more profound comprehension of the late cosmic acceleration \cite{Ferraro08,Geng11,Cai16,Jarv16}. Additionally, it is worth noting that the metric theories can be generalized beyond Riemannian geometry, as demonstrated in \cite{Conroy18}. Building upon the metric-affine formalism and an extension of symmetric teleparallel gravity, Harko et al. \cite{Harko18} proposed the modified $f(Q)$ gravity.

Recent literature on extended theories of gravity places significant importance on investigating the astrophysical and cosmological properties of $f(Q)$ gravity. In this context, Hohmann et al. \cite{Hohmann19} examined the propagation velocity and potential polarizations of gravitational waves taking Minkowski spacetime in $f(Q)$ gravity. Soudi et al. \cite{Soudi19} constrained the strong-field behavior of this gravitational theory by analyzing gravitational wave polarizations. Numerous studies have explored $f(Q)$ gravity in various backgrounds, including the consideration of observational data constraints for different parametrizations of $f(Q)$ \cite{Lazkoz19,Ayuso21}, energy conditions \cite{Mandal20}, cosmography \cite{Mandal20a}, bouncing scenarios \cite{Mandal21,Bajardi20}, black holes \cite{D'Ambrosio22}, and the evolution of the growth index in matter perturbations, among others \cite{Khyllep21}. Further, comprehensive discussions on the $f(Q)$ theory can be found in recent references such as \cite{Esposito22,Zhao22,Albuquerque22,Atayde21,Frusciante21,Dimakis21}. It is noteworthy to highlight that $f(Q)$ gravity effectively explains the accelerated expansion of the universe with a statistical precision comparable to other established modified theories of gravity. Consequently, modified gravities at an astrophysical scale should be considered to overcome this so-called degeneracy. Promising techniques for this objective encompass direct observation of the galactic core \cite{Akiyama19} and the growing tally of detections of gravitational waves \cite{Abbott16}. For a more detailed exploration of this topic, the following references \cite{ae1,ae2,ae3} provide contemporary works in various contexts. 

The process of stellar evolution encompasses the continuous changes in the internal composition of a star from its formation to its eventual demise. Throughout this transformative journey, a star emits a substantial amount of energy in the form of photons or neutrinos. The enigmatic characteristics of these celestial objects have motivated numerous researchers to explore their relevance in the realms of cosmology and astrophysics. Gravitational collapse stands out as a pivotal subject in relativistic astrophysics, closely connected to the genesis of self-gravitating structures. Initially, a stable system maintains a state of hydrostatic equilibrium until its own gravitational forces induce a gradual reduction in size, culminating in gravitational collapse. The outcome of this collapse results in various structures such as white dwarfs, neutron stars, or black holes (collectively known as compact stars), depending upon the initial mass of the star. The composition and nature of compact stars have intrigued physicists, prompting investigations into their diverse evolutionary stages and internal characteristics in the field of gravitational physics. Numerous attempts have been made to investigate the characteristics of stellar models under various physical effects. In this context, the notion of anisotropy plays a crucial role in studying the structural mechanisms of compact systems \cite{a1}. Bower and Liang \cite{b} conducted an analysis of locally anisotropic pressure distributions in spherically symmetric matter, demonstrating that pressure anisotropy significantly influences the parameters governing the hydrostatic equilibrium of celestial systems. Maurya and his team \cite{c} examined compact spherically symmetric systems, considering anisotropic pressure profiles, and their findings provide valuable insights into assessing the stability of compact objects. Karamakar and collaborators \cite{d} studied the role of anisotropy in cold compact stars.

Many researchers have shown a keen interest in exploring the intrinsic and extrinsic aspects of stellar bodies by applying gravitational potentials. One notable approach is the Karmarkar condition, which establishes a relationship between temporal and radial metric components, facilitating the study of compact objects. The Karmakar condition involves embedding a four-dimensional Riemann manifold into a higher-dimensional manifold. This technique is used to analyze the solutions of static spherical spacetime and discuss the physically viable characteristics of anisotropic stars \cite{e,f}. Bhar et al. \cite{g} applied this approach to investigate the stability conditions of anisotropic compact stars. Makalo et al. \cite{h} studied exact solutions for charged stars using the Karmarkar condition and found that interaction between potentials and charge leads to the formulation of a realistic anisotropic model. Sharif and Gul \cite{i} explored physically viable and stable anisotropic stellar models in the vicinity of the Karmarkar approach. The physical properties of stellar objects have also been discussed in different modified theories of gravity using the Karmarkar condition \cite{j,k,l}. Spherically symmetric solutions are explored in great detail within the framework of $f(R)$ gravity (see, for instance, especially Refs. \cite{Vilja06,Dobado09,Kobayashi08,Upadhye09,Manzoor22}). Under the $f(T)$ framework, the same symmetry is also investigated (see, e.g., Refs. \cite{Aftergood14,DeBenedictis16,Lin17,Krssak15}). The dynamical aspects of compact stars have also been studied in other modified theories of gravity like $f(G)$ \cite{Abbas15,Malik22}, $f(Q)$ \cite{Mandal22,Errehymy22}, $f(G,T)$ \cite{Shamir17,Sadiq22}, etc. One can find many other studies on compact stars in the vicinity of different modified theories and scenarios \cite{1,2,3,4,5,6,7,8}.

Exact solutions are of great significance, especially for spherically symmetric fluid distributions in astrophysical compact objects. Numerous authors have imposed different constraints on the spacetime geometry and matter distribution to derive exact solutions of the Einstein field equations. Delgaty and Lake \cite{m} introduced elementary criteria to assess the physical viability of these exact solutions. While formulating several static spherically symmetric exact solutions, they emphasized that only a subset of these solutions is physically meaningful. Correcting the solution provided by Duorah and Ray \cite{n}, Finch and Skea \cite{o} derived a new exact solution to the Einstein field equations. An intriguing aspect of their model is that it describes a compact star with isotropic pressure. The Finch-Skea metric \cite{o} is indeed well-behaved, meeting all the criteria to be considered as a Delgaty and Lake solution \cite{m} for a static spherically symmetric perfect fluid model. The intriguing aspect of the Finch-Skea metric lies in the formulation of a radial potential that enables the comprehensive integration of the field equations \cite{p}. This, in turn, facilitates the determination of all remaining geometric and dynamical quantities. Naeem et al. \cite{q} devised a new physical generalized polytropic model featuring anisotropic matter distribution in the context of Finch–Skea approach. Bhar et al. \cite{r} solved the field equations by using Finch–Skea ansatz and analyzed stable spherically symmetric spacetime in the context of $f(R,T)$ gravity. Sokoliuk et al. \cite{s} studied anisotropic stellar solutions that embrace the Finch-Skea symmetry, characterized by viable and non-singular metric potentials, in the presence of exotic matter fields. Das et al. \cite{t} presented a class of exact solutions for relativistic stellar systems by taking Finch–Skea ansatz with and without anisotropy.

This paper aims to explore the physical properties of spherically symmetric configurations with anisotropic distributions in the framework of $f(Q)$ symmetric teleparallel gravity. The organization of the article is as follows: Section \ref{sec1} provides a brief introduction, while Section \ref{sec2} presents a comprehensive review of the field equations in $f(Q)$ gravity and the equations of motion for a charged fluid sphere. In Section \ref{sec3}, we generalize the dynamics and geometry governing static spherically symmetric configurations using the Karmarker condition. The matching of the anisotropic interior solution in $f(Q)$ gravity to the exterior Reissner-Nordstrom vacuum solution is discussed in Section \ref{sec4}. Section \ref{sec5} focuses on the physical analysis of the compact star solutions obtained in this study. Finally, concluding remarks are provided to summarize the findings of this article. Throughout the study, natural (geometrized) units are employed, where $G$ and $c$ are set to 1, and the metric signature is given by $\left(-,+,+,~+\right)$.

\section{General Formalism of $f(Q)$ Gravity and the Field Equations}\label{sec2}
In this section, we provide basic formalism of $f(Q)$ gravity described by the action \cite{Jimenez18}, expressed as
\begin{equation}\label{9}
S=\int\sqrt{-g}d^{4}x\Big[\frac{1}{2}f(Q)+\lambda_{\alpha}^{\beta\mu\nu}R^{\alpha}_{\beta\mu\nu}+\lambda{\alpha}^{\mu\nu}T^{\alpha}{\mu\nu}+L{m}\Big],
\end{equation}
where $f(Q)$ is a function of non-metricity $Q$, $\lambda_{\alpha}^{\beta\mu\nu}$ correspond to the multipliers for the Lagrangian while $L_{m}$ represents the Lagrangian density of matter distribution. Non-metricity in the framework of the affine connection can be defined as follows
\begin{equation}\label{1}
Q_{\alpha \epsilon \upsilon} \equiv \nabla_{\alpha} g_{\epsilon \upsilon}=\partial_{\alpha} g_{\epsilon \upsilon}-\Gamma_{\alpha \epsilon}^{\varrho} g_{\varrho \upsilon}-\Gamma_{\alpha \upsilon}^{\varrho} g_{\epsilon \varrho}.
\end{equation}
The standard form of the affine connection can be split into three components given by
\begin{equation}\label{2}
\Gamma_{\epsilon \upsilon}^{\varrho}=\left\{{ }_{\epsilon \upsilon}^{\varrho}\right\}+K_{\epsilon \upsilon}^{\varrho}+L_{\epsilon \upsilon}^{\varrho}.
\end{equation}
In the above equation, the expression $\left\{{ }_{\epsilon \upsilon}^{\varrho}\right\}$ corresponds to the Levi-Civita connection that can be defined by the metric $g_{\epsilon \upsilon}$ as
\begin{equation}\label{3}
\left\{{ }_{\epsilon \upsilon}^{\varrho}\right\} \equiv \frac{1}{2} g^{\varrho \beta}\left(\partial_{\epsilon} g_{\beta \upsilon}+\partial_{\upsilon} g_{\beta \epsilon}-\partial_{\beta} g_{\epsilon \upsilon}\right),
\end{equation}
where $K_{\epsilon \upsilon}^{\varrho}$ refers to the contortion defined by the relation
\begin{equation}\label{4}
K_{\epsilon \upsilon}^{\varrho} \equiv \frac{1}{2} T_{\epsilon \upsilon}^{\varrho}+T{}_({_\epsilon}{}^{\varrho}{}_{\upsilon)}.
\end{equation}
Here, torsion tensor $T_{\epsilon \upsilon}^{\varrho}$ is recognized as the antisymmetric part of the affine connection $T_{\epsilon \upsilon}^{\varrho} \equiv 2 \Gamma_{[\epsilon \upsilon]}^{\varrho}$ whereas the disformation $L_{\epsilon \upsilon}^{\varrho}$ can be expressed as
\begin{equation}\label{5}
L_{\epsilon \upsilon}^{\varrho} \equiv \frac{1}{2} Q_{\epsilon \upsilon}^{\varrho}-Q{}_({_\epsilon}{}^{\varrho}{}_{\upsilon)}.
\end{equation}
Now, we calculate the non-metricity conjugate as follows
\begin{equation}\label{6}
P_{\epsilon \upsilon}^{\alpha}=-\frac{1}{4} Q_{\epsilon \upsilon}^{\alpha}+\frac{1}{2} Q_{\left(\epsilon^{\alpha} \upsilon\right)}^{\alpha}+\frac{1}{4}\left(Q^{\alpha}-\tilde{Q}^{\alpha}\right) g_{\epsilon \upsilon}-\frac{1}{4} \delta^{\alpha}{ }_{(\epsilon} Q_{\upsilon)},
\end{equation}
with two different independent traces
\begin{equation}\label{7}
Q_{\alpha} \equiv Q_{\alpha}{ }^{\epsilon}{ }_{\epsilon}, \quad \tilde{Q}_{\alpha} \equiv Q_{\alpha \epsilon}^{\epsilon}.
\end{equation}
Consequently, the non-metricity scalar yields
\begin{equation}\label{8}
Q=-Q_{\alpha \epsilon \upsilon} P^{\alpha \epsilon \upsilon}.
\end{equation}
Varying the action in Eq.(\ref{9}) with respect to $g_{\epsilon \upsilon}$, we derive the following field equations
\begin{eqnarray}\label{10}
-T_{\epsilon \upsilon}&=& \frac{2}{\sqrt{-g}} \nabla_{\alpha}\left(\sqrt{-g} f_{Q} P_{\epsilon \upsilon}^{\alpha}\right)+\frac{1}{2} g_{\epsilon \upsilon} f \nonumber\\&+&f_{Q}\left(P_{\epsilon \alpha \beta} Q_{\upsilon}{ }^{\alpha \beta}-2 Q_{\alpha \beta \epsilon} P^{\alpha \beta}{ }_{\upsilon}\right),
\end{eqnarray}
where $ f_{Q} \equiv \partial_{Q} f(Q)$. The standard form of the energy-momentum tensor is given by
\begin{equation}\label{11}
T_{\epsilon \upsilon} \equiv-\frac{2}{\sqrt{-g}} \frac{\delta\left(\sqrt{-g} L_{m}\right)}{\delta g^{\epsilon \upsilon}}.
\end{equation}
Furthermore, varying Eq.(\ref{9}) with respect to the connection, we get
\begin{equation}\label{12}
\nabla_{\rho} \varrho_{\alpha}{ }^{\upsilon \epsilon \rho}+\varrho_{\alpha}{ }^{\epsilon \upsilon}=\sqrt{-g} f_{Q} P_{\epsilon \upsilon}^{\alpha}+H_{\alpha}{ }^{\epsilon \upsilon}.
\end{equation}
The density for the hyper-momentum tensor has the form
\begin{equation}\label{13}
H_{\alpha}{ }^{\epsilon \upsilon}=-\frac{1}{2} \frac{\delta L_{m}}{\delta \Gamma^{\alpha}{ }_{\epsilon \upsilon}}.
\end{equation}
Considering the antisymmetric property of $\upsilon$ and $\epsilon$ in the Lagrangian multiplier coefficients, Eq.(\ref{12}) can be reduced to the following relation
\begin{equation}\label{14}
\nabla_{\epsilon} \nabla_{\upsilon}\left(\sqrt{-g} f_{Q} P^{\epsilon \upsilon}{ }_{\alpha}+H_{\alpha}{ }^{\epsilon \upsilon}\right)=0.
\end{equation}
Using $\nabla_{\epsilon} \nabla_{\upsilon} H_{\alpha}{ }^{\epsilon \upsilon}=0$, we have
\begin{equation}\label{15}
\nabla_{\epsilon} \nabla_{\upsilon}\left(\sqrt{-g} f_{Q} P^{\epsilon \upsilon}\right)=0.
\end{equation}
Also, the affine connection, without torsion and curvature, has the following form
\begin{equation}\label{16}
\Gamma_{\epsilon \upsilon}^{\alpha}=\left(\frac{\partial x^{\alpha}}{\partial \xi^{\varrho}}\right) \partial_{\epsilon} \partial_{\upsilon} \xi^{\varrho}.
\end{equation}
Taking a special coordinate choice, in the case of coincident gauge, the non-metricity reduces to the following expression
\begin{equation}\label{170}
Q_{\alpha \epsilon \upsilon}=\partial_{\alpha} g_{\epsilon \upsilon}.
\end{equation}
In the context of $f(Q)$ gravity, the field equations are formulated to ensure the conservation of the energy-momentum tensor. To investigate stellar configurations, we primarily focus on finding the gravitational field equations that govern static and spherically symmetric spacetimes defined by Eq.(\ref{10}).

To describe the structural properties of stellar objects in $f(Q)$ gravity, we choose a spherically symmetric line element 
\begin{equation}\label{180}
ds^{2}=-e^{\nu(r)}dt^2+e^{\lambda(r)}dr^2+r^2\sin^{2}\theta d\phi^{2}+d\theta^{2},
\end{equation}
where $\nu(r)$ and $\lambda(r)$ are the functions of radial coordinate $r$. Substituting Eq.(\ref{180}) into (\ref{8}), we obtain the non-metricity scalar $Q$ as
\begin{equation}\label{190}
Q=-\frac{\left(2 e^{-\lambda (r)}\right) \left(\nu '(r)+\frac{1}{r}\right)}{r},
\end{equation}
where prime stands for the derivative with respect to $r$.

The energy-momentum tensor for a charged anisotropic matter is given by
\begin{equation}\label{20}
T_{ij}=(\rho+p_{t})\varsigma_{i}\varsigma_{j}-p_{t}g_{ij}+(p_{r}-p_{t})\xi_{i}\xi_{j},
\end{equation}
where $\rho, p_{r},$ and $p_{t}$ represent the energy density, radial pressure, and tangential pressure components, respectively.
The radial four-vector $\xi_{\alpha}$ and the four-velocity  $\varsigma_{i}$ are introduced to meet the following requirement
\begin{equation*}
\varsigma^{\alpha}=e^{\frac{-\nu}{2}}\delta^{\alpha}_{0},~~~\varsigma^{\alpha}\varsigma_{\alpha}=1,~~~\xi^{\alpha}=e^{\frac{-\lambda}{2}}\delta^{\alpha}_{1},
~~~\xi^{\alpha}\xi_{\alpha}=-1.
\end{equation*}
Inducing the metric (\ref{18}) and the anisotropic matter distribution (\ref{20}), we can extract the nonzero components of the field equations from the equations of motion (\ref{10}) as
\begin{eqnarray}
\rho &=& -f_Q \left(Q+\frac{1}{r^2}+\frac{e^{-\lambda (r)} \left(\lambda '(r)+\nu '(r)\right)}{r}\right)+\frac{f}{2},~~~~\label{26}\\
p_r &=& f_Q \left(Q+\frac{1}{r^2}\right)-\frac{f}{2},~\label{27}\\
p_t &=& f_Q \bigg(\frac{Q}{2}-e^{-\lambda (r)} \Bigg(\Bigg(\frac{\nu '(r)}{4}+\frac{1}{2 r}\Bigg) \left(\nu '(r)-\lambda '(r)\right)\nonumber\\ &+& \frac{\nu ''(r)}{2}\Bigg)\Bigg)-\frac{f}{2},\label{28}\\
0&=&\frac{cot\theta}{2}Q^{'}f_{QQ}.\label{29}
\end{eqnarray}

 The linear form of $f(Q)$ gravity is obtained through Eq.(\ref{29}) as
\begin{equation}\label{32}
f=\alpha\,  Q+\beta,
\end{equation}
where the $\beta$ symbolizes an integration constant. 
Eisenhart \cite{51} demonstrated that a class-I $(n+1)$-dimensional space $V^{n+1}$ can be embedded into an $(n+2)$-dimensional Euclidean space $E^{n+2}$ and provided a symmetric tensor that satisfies the Gauss-Codazzi equations
\begin{eqnarray*}R_{stpq}&=&2ea_{s[p{}a_{q}]t}, \\ \qquad a_{s[t;p]}-\Gamma^{q}{}_{[tp]}a_{sq}+\Gamma^{q}_{s[t{}a_{p}]q}&=&0,
\end{eqnarray*}

where $R_{stpq}$ is the Riemannian curvature tensor, $e=\pm1$ and the square brackets indicate anti-symmetrization. One can easily find one of the most convenient methods in the literature for the calculation of the components $e^{\nu(r)}$ and $e^{\lambda(r)}$ of a spherically symmetric spacetime by using the Karmarkar condition. This condition provides a necessary and sufficient condition for the embedding of a class-I metric. The Karmarkar condition is defined as 
\begin{equation}\label{150}
R_{0101}R_{2323}=R_{0202}R_{1313}-R_{1202}R_{1303}.
\end{equation}
The non-vanishing components of Riemannian tensor for metric (\ref{3}) are given by
\begin{eqnarray*}
R_{0101}&=&-\frac{1}{4}e^{\nu(r)}\left(-a'(r)\lambda'(r)+\nu'^{2}(r)+2\nu''(r)\right),\quad\\
R_{2323}&=&-r^{2}\sin^{2}\theta\left(1-e^{-\lambda(r)}\right),\quad\\
R_{0202}&=&-\frac{1}{2}r\nu'(r)e^{\nu(r)-\lambda(r)},\\
R_{1313}&=&-\frac{1}{2}\lambda'(r) r\sin^{2}\theta,\quad
R_{1202}=0,\quad
R_{1303}=0.
\end{eqnarray*}
Inserting these components in Eq.(\ref{150}), we find the following
differential equation 
\begin{equation}\label{160}
\left(\lambda'(r)-\nu'(r)\right) \nu'(r)e^{\lambda(r)}+ 2\left(1-e^{\lambda(r)}\right)\nu''(r)+\nu'^2(r)=0.
\end{equation}
It is interesting to mention here that the embedding class-I solutions can be obtained by Eq.(\ref{160}) as they can be embedded in a 5-dimensional Euclidean Space.
Integrating Eq.(\ref{160}) for $e^a$, we get its expression in terms of $e^b$ as follows
\begin{equation}\label{17}
e^{\nu(r)}=\left(A+B \int \sqrt{e^{\lambda(r)}-1} \, dr\right)^2,
\end{equation}
where $A$ and $B$ are the constants of integration.
Consequently, the line element (\ref{3}) is given by
\begin{eqnarray}\label{17aaa}
ds^{2}&=&\left(A+B \int \sqrt{e^{\lambda(r)}-1} \, dr\right)^2dt^{2}-e^{\lambda(r)} dr^{2}\nonumber\\&-&r^{2} d\theta^{2}-r^{2}\sin^{2}\theta d\phi^{2}.
\end{eqnarray}
Eqs. (\ref{17}) and (\ref{17aaa}) contain $\lambda(r)$, which is the function of $r$, so both of these equations (i.e. Eqs. (\ref{17}) and (\ref{17aaa})) are dependent on $r$. Thus, we conclude that the spherical symmetric spacetime depends only on the generating function $e^{\lambda(r)}$.

\section{Generalized Solution for a Compact Star}\label{sec3}

In this section, we derive expressions for various quantities such as density, radial and tangential pressures, etc., by assuming a specific viable model for the metric component $e^{\lambda(r)}$. From Eq.(\ref{17}), it is observed that the metric potentials $e^{\nu(r)}$ and $e^{\lambda(r)}$ are related to each other which makes it convenient to evaluate one of the metric components through this relation if the other one is known. To construct a realistic anisotropic model, we consider a well-known model \cite{52} for $g_{rr}$ given by
\begin{equation}\label{18}
\lambda(r)=\log \left(1+\frac{c r^2 \left(a r^2+1\right)^n}{\left(b r^2+1\right)^2}\right),
\end{equation}
where $a$, $b$, $c$, and $n$ are the arbitrary constants. It is evident that the solution presented in Eq.(\ref{18}) is regular, accurate, and provides a reasonable approximation of the neutron star model for a specific range of parameters associated with $g_{rr}$ \cite{52}. Substituting Eq.(\ref{18}) into (\ref{17}), we have
\begin{widetext}
\begin{equation}\label{19}
e^{\nu(r)}=\log \left(A+\frac{B \left(b r^2+1\right) \left(\frac{a b r^2+b}{a b r^2+a}\right)^{-\frac{n}{2}} \sqrt{\frac{c r^2 \left(a r^2+1\right)^n}{\left(b r^2+1\right)^2}} \, _2F_1\left(-\frac{n}{2},-\frac{n}{2};1-\frac{n}{2};\frac{a-b}{a b r^2+a}\right)}{b n r}\right)^{2}.
\end{equation}
\end{widetext}
Using the value of metric potentials $\nu(r)$ and $\lambda(r)$ from Eqs.(\ref{18}) and (\ref{19}) along with the torsion scalar
and its derivative, the expressions for $\rho$, $p_{r}$, $p_{t}$ and $\vartriangle$ turn out to be
\begin{widetext}
\begin{eqnarray}\nonumber
\rho&=&-\frac{2 \alpha  c l_2 r \left(b r^2+1\right) \left(a r^2+1\right)^{n-1}}{\left(c r^2 \left(a r^2+1\right)^n+b^2 r^4+2 b r^2+1\right)^2}-\frac{\alpha  c \left(a r^2+1\right)^n}{c r^2 \left(a r^2+1\right)^n+b^2 r^4+2 b r^2+1}+\frac{\beta }{2},\\\nonumber
p_{r} &=&\alpha  c \left(a r^2+1\right)^n \left(\frac{1}{c r^2 \left(a r^2+1\right)^n+b^2 r^4+2 b r^2+1}-\frac{2 b B n r \left(\frac{a b r^2+b}{a b r^2+a}\right)^{n/2}}{\left(b r^2+1\right) \left(l_1 r\right) \left(\frac{c r^2 \left(a r^2+1\right)^n}{\left(b r^2+1\right)^2}+1\right)}\right)-\frac{\beta }{2},\\\nonumber
p_{t}&=&-\frac{1}{4 \left(\frac{c r^3 \left(a r^2+1\right)^n}{\left(b r^2+1\right)^2}+r\right)}\Big[\frac{4 \alpha  b^2 B^2 c^2 n^2 r^5 \left(a r^2+1\right)^{2 n} \left(\frac{a b r^2+b}{a b r^2+a}\right)^n}{\left(b r^2+1\right)^2 \left(l_1 r\right){}^2}+\frac{4 \alpha  b B c n r^2 \left(a r^2+1\right)^n \left(\frac{a b r^2+b}{a b r^2+a}\right)^{n/2}}{\left(b r^2+1\right) \left(l_1 r\right)}\nonumber\\&+&\frac{4 \text{$\alpha $l}_3 b B n r \left(b r^2+1\right) \left(\frac{a b r^2+b}{a b r^2+a}\right)^{n/2} \left(\frac{c r^2 \left(a r^2+1\right)^n}{\left(b r^2+1\right)^2}\right)^{3/2}}{\left(a r^3+r\right) \left(l_1 r\right){}^2}+2 r \beta  \left(\frac{c r^2 \left(a r^2+1\right)^n}{\left(b r^2+1\right)^2}+1\right)-l_4 r\big],\\
\Delta &=&p_t-p_r,
\end{eqnarray}
\begin{eqnarray}\nonumber
l_1 (r)&=&A b n \left(b r^2+1\right) \left(\frac{a b r^2+b}{a b r^2+a}\right)^{n/2} \sqrt{\frac{c r^2 \left(a r^2+1\right)^n}{\left(b r^2+1\right)^2}}+B c r \left(a r^2+1\right)^n \, _2F_1\left(-\frac{n}{2},-\frac{n}{2};1-\frac{n}{2};\frac{a-b}{a b r^2+a}\right),\\
l_2 (r) &=& (1 - b r^2 + a r^2 (1 + n - b r^2 + b n r^2)),\\
l_3 (r)&=&b n r \left(\frac{a b r^2+b}{a b r^2+a}\right)^{n/2} \Big[a A r^2 \left(b n r^2-b r^2+n+1\right)-B r \left(a r^2+1\right) \left(b r^2+1\right) \sqrt{\frac{c r^2 \left(a r^2+1\right)^n}{\left(b r^2+1\right)^2}}-A b r^2+A\Big]\\&+&B l_2 r \left(b r^2+1\right) \sqrt{\frac{c r^2 \left(a r^2+1\right)^n}{\left(b r^2+1\right)^2}} \, _2F_1\left(-\frac{n}{2},-\frac{n}{2};1-\frac{n}{2};\frac{a-b}{a b r^2+a}\right),~~~~
\end{eqnarray}
\begin{eqnarray}
l_4 (r)&=&\frac{2 \alpha  c l_2 r r \left(a r^2+1\right)^{n-1} \left(\frac{2 b B c n r^3 \left(a r^2+1\right)^n \left(\frac{a b r^2+b}{a b r^2+a}\right)^{n/2}}{\left(b r^2+1\right) \left(l_1 r\right)}+2\right)}{\left(b r^2+1\right)^3 \left(\frac{c r^2 \left(a r^2+1\right)^n}{\left(b r^2+1\right)^2}+1\right)}.
\end{eqnarray}
\end{widetext}

\section{Calculation of the Model Constants Using Boundary Conditions}\label{sec4}

This section is devoted to finding the values of model constants by matching the interior and exterior geometries.
In this context, the comparison of the interior spacetime (\ref{3}) with the exterior Schwarzschild metric yields, 

\begin{eqnarray}\label{23}
ds^2=\left(1-\frac{2M}{r}-\frac{\Lambda}{3} r^{2}\right)dt^2-\left(1-\frac{2M}{r}-\frac{\Lambda}{3}r^{2}\right)^{-1}dr^2\nonumber\\-r^2\left(d \theta^2 + \sin^2 \theta d\phi^2\right),
\end{eqnarray}
where $\Lambda$ denotes a cosmological constant with $\Lambda= \beta/2\alpha$. It is noticed that when $\beta = 0$, the Reissner–Nordström (RN) de-Sitter (dS) spacetime reduces to RN exterior solution. 
We take $r=R$ at the boundary. Comparing Eqs.(\ref{3}) and (\ref{23}), we get
\begin{equation}\label{24}
1-\frac{2M}{R}=e^{\nu(R)}=e^{-\lambda(R)},
\end{equation}
\begin{equation}\label{25}
P_{r}(r=R)=0.
\end{equation}
Substituting $e^{\nu(r)}$ and $e^{\lambda(r)}$ in the boundary conditions (\ref{24}) and (\ref{25}), we evaluate the following constants
\begin{widetext}
\begin{eqnarray}\label{25*}
A&=&\sqrt{1-\frac{2 M}{R}}-\frac{B \left(b R^2+1\right) \left(\frac{a b R^2+b}{a b R^2+a}\right)^{-\frac{n}{2}} \sqrt{\frac{c R^2 \left(a R^2+1\right)^n}{\left(b R^2+1\right)^2}} \, _2F_1\left(-\frac{n}{2},-\frac{n}{2};1-\frac{n}{2};\frac{a-b}{a b R^2+a}\right)}{b n R},\\\nonumber
B&=&\frac{R \sqrt{1-\frac{2 M}{R}} \left(c \left(a R^2+1\right)^n \left(2 \alpha -R^2\beta \right)-\beta \left(b R^2+1\right)^2\right)}{4 \alpha  \left(b R^2+1\right)^2 \sqrt{\frac{c R^2 \left(a R^2+1\right)^n}{\left(b R^2+1\right)^2}}},\\\nonumber
c&=&-\frac{2 M \left(b R^2+1\right)^2 \left(a R^2+1\right)^{-n}}{R^2 (2 M-R)}.
\end{eqnarray}
\end{widetext}
Here $M$ and $R$ represent the mass and radius of the compact star, respectively while $\alpha$, $\beta,~d$, and $M$ are the free parameters. Since the observed cosmological constant value in the universe is of order $10^{-34}/km^2$. Therefore, its effect on the compact stars study will be negligible. Therefore, we can assume it value to be zero which leads $\beta=0$. 

\begin{table}
\caption{\label{tab1}{Model parameters using mass and radius of compact star \textbf{PSR J0437-4715 ($M_0=1.76$ and $R=10.5$ km)} for different values of $n$ by fixing $\alpha= -2.3,\;a=0.03,\;\&\;b=0.0015$.}}
\label{Table1}
\begin{center}
\vspace{0.48cm}\begin{tabular}{|c|c|c|c|c|c|c|c|c|c|c|}
\hline
    n & A         & B        & c                        & $\frac{p_{rc}}{\rho_c}$ ($r=0$) \\
\hline
 1.0 & -1.53033 & 0.0275713 & 0.00466814       &  $<1$\\
\hline
 1.2& -1.54533 &  0.0275713& 0.00440884                 & $<1$\\
\hline
   1.4& -1.80812 & 0.0275713 & 0.00416394                    &$<1$\\
\hline
  1.5& -2.09165 & 0.0275713 & 0.00404665                   &$<1$\\
\hline
 1.6 & -2.5643 & 0.0275713 & 0.00393265            & $<1$\\
\hline
   1.7& -3.40587 &  0.0275713& 0.00382187                    &$<1$\\
\hline
    1.8& -5.15935 & 0.0275713 & 0.00371421                    &$<1$\\
\hline
   1.9& -10.5453 & 0.0275713 & 0.00360958                  & $<1$\\
\hline
\end{tabular}\\
\vspace{0.5cm}
\begin{tabular}{|c|c|c|c|c|c|c|c|c|c|c|}
\hline
    Expression          & Remarks  \\
\hline
     $\rho$ ,  $p_r$ ,  $p_t$             & $>0$, acceptable          \\
\hline
 $\frac{p_{rc}}{\rho_c}$  &  $\leq1$                     \\
\hline
 $\Delta$               & $>0$, acceptable          \\
\hline
$\frac{d \rho}{dr}$, $\frac{d p_r}{dr}$,
$\frac{d p_t}{dr}$      &  $<0$, acceptable       \\
\hline
$F_a$, $F_h$, and $F_g$ & balanced                  \\
\hline
 $\rho\pm p_r$             & $>0$, acceptable          \\
\hline
 $\rho\pm p_t$            & $>0$, acceptable           \\
\hline
$\rho+p_r+2 p_t$       & $>0$, acceptable          \\
\hline
$m(r)$                & $>0$, acceptable            \\
\hline
$u(r)$                & $0<u(r)<\frac{8}{9}$, acceptable  \\
\hline
$z_s$                 & $0<z_s)<5$, acceptable       \\
\hline
$w_r$ , $w_t$                 & $0<w_r<1$, acceptable        \\
\hline
$v_{r}^{2}$  ,$v_{t}^{2}$        &  $0<v_{r}^{2}<1$, acceptable \\
\hline
$v_{t}^{2}-v_{r}^{2}$ & $-1<\mid v_{t}^{2}-v_{r}^{2}\mid<1$, acceptable \\
\hline
$\Gamma_r$          &   $>\frac{4}{3}$, acceptable   \\
\hline
\end{tabular}
\end{center}
\end{table}

\section{Graphical Illustration of Physical Characteristics of the Constructed Model}\label{sec5}

Now we illustrate the stellar structure visually in the framework of $f(Q)$ theory of gravity. To ensure a realistic and viable stellar configuration, we explore some important physical characteristics and constraints that must be satisfied. In the following, we will present graphical representations of our compact star structure, including the metric potential, energy-density, pressure profiles, pressure gradients with respect to the radial coordinate, Zeldovich's condition, stability, causality, and adiabatic index. To analyze these features, we will consider the mass and radii of three astronomical objects, namely $PSR J0437-4715$. It is important to list the fundamental properties satisfied by a physically valid stellar configuration.
\begin{itemize}
\item Matching of the interior and exterior spacetime at the boundary.
\item The metric components should exhibit positive regular behavior, especially at the center, with a monotonically increasing trend.
\item The profiles of energy density and pressures should be positive, regular, and monotonically decreasing, with the maximum occurring at the center.
\item The anisotropic parameter $\Delta$ should be greater than zero within the star boundary.
\item The energy constraints must be satisfied by the stellar structure.
\item For an interior region of the star, the radial and tangential speeds of sound should be less than the speed of light, adhering to the causality condition.
\item The stellar structure should be stable, assessed through the adiabatic index and the Tolman-Oppenheimer-Volkoff (TOV) equation.
\item The mass function should be free from central singularities, and the surface redshift and compactness parameter should exhibit physically reasonable behavior.
\item The density and pressure profiles at the center should satisfy the well-known Zeldovich's condition.
\end{itemize}

\begin{figure}
\centering \includegraphics[width=70mm,height=50mm]{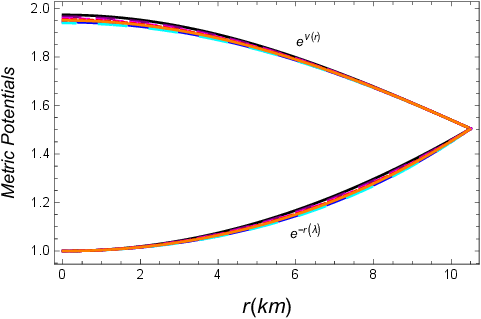}
\includegraphics[width=70mm,height=50mm]{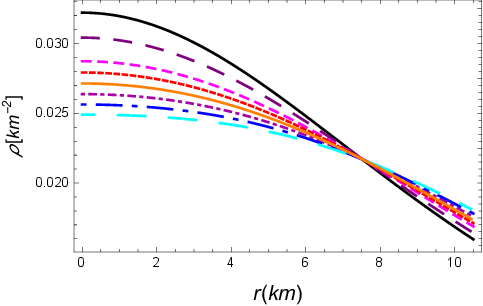}
\caption{Behavior of metric potentials and energy density versus radial coordinate $r$. Here $n=1.0$ (Black with maximum density) and $n=1.9$ (Cyan with minimum density). The specific values of constant parameters are mentioned in Table-\ref{tab1}.\{$n=1 (\textcolor{black}{\bigstar})$, $n=1.2(\textcolor{purple}{\bigstar})$, $n=1.4\;(\textcolor{magenta}{\bigstar})$, $n=1.5(\textcolor{red}{\bigstar})$, $n=1.6(\textcolor{orange}{\bigstar})$, $n=1.7(\textcolor{violet}{\bigstar})$, $n=1.8\;(\textcolor{blue}{\bigstar})$, and $n=1.9(\textcolor{cyan}{\bigstar})$.\}}
\end{figure}\label{fig1}

\begin{figure}
\includegraphics[width=70mm,height=50mm]{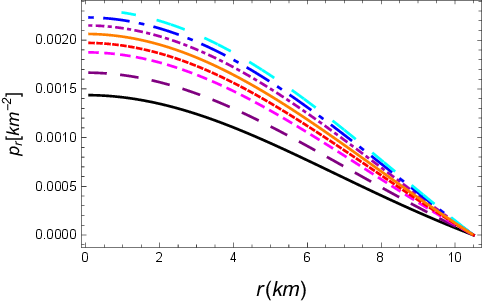}
\includegraphics[width=70mm,height=50mm]{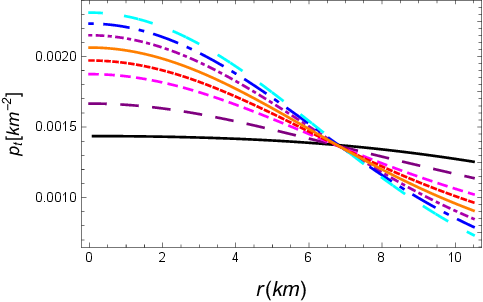}
\caption{Behavior of radial and tangential pressure versus radial coordinate $r$. Here $n=1.0$ (Black with maximum density) and $n=1.9$ (Cyan with minimum density). The specific values of constant parameters are mentioned in Table-\ref{tab1}.\{$n=1 (\textcolor{black}{\bigstar})$, $n=1.2(\textcolor{purple}{\bigstar})$, $n=1.4\;(\textcolor{magenta}{\bigstar})$, $n=1.5(\textcolor{red}{\bigstar})$, $n=1.6(\textcolor{orange}{\bigstar})$, $n=1.7(\textcolor{violet}{\bigstar})$, $n=1.8\;(\textcolor{blue}{\bigstar})$, and $n=1.9(\textcolor{cyan}{\bigstar})$.\}}
\end{figure}\label{fig2}

\begin{figure}
\includegraphics[width=70mm,height=50mm]{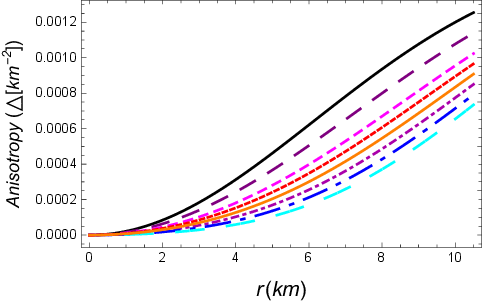}\\
\includegraphics[width=70mm,height=50mm]{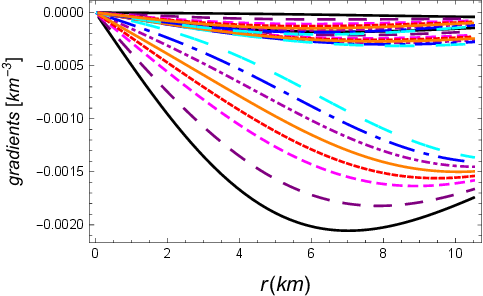}\\\includegraphics[width=70mm,height=50mm]{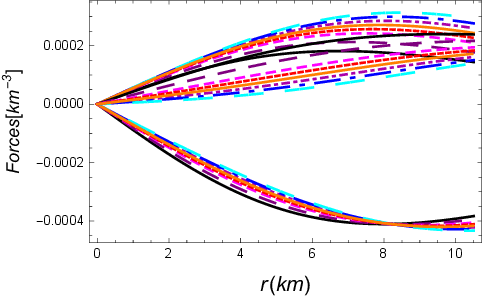}
\caption{Behaviors of anisotropy, gradients, and forces versus radial coordinate $r$. Here $n=1.0$ (Black with maximum density) and $n=1.9$ (Cyan with minimum density). The specific values of constant parameters are mentioned in Table-\ref{tab1}.\{$n=1 (\textcolor{black}{\bigstar})$, $n=1.2(\textcolor{purple}{\bigstar})$, $n=1.4\;(\textcolor{magenta}{\bigstar})$, $n=1.5(\textcolor{red}{\bigstar})$, $n=1.6(\textcolor{orange}{\bigstar})$, $n=1.7(\textcolor{violet}{\bigstar})$, $n=1.8\;(\textcolor{blue}{\bigstar})$, and $n=1.9(\textcolor{cyan}{\bigstar})$.\}}
\end{figure}\label{fig3}

\begin{figure}
\includegraphics[width=70mm,height=50mm]{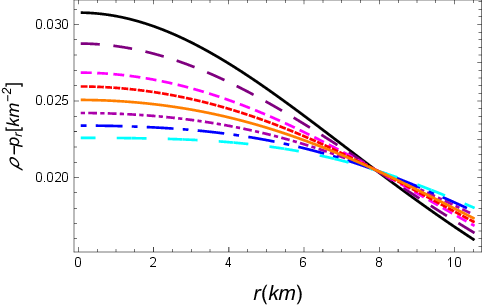}\space\includegraphics[width=70mm,height=50mm]{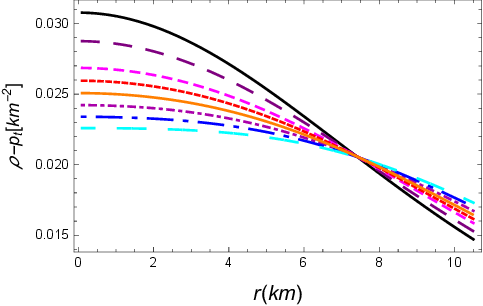}
\caption{Energy conditions versus radial coordinate $r$. Here $n=1.0$ (Black with maximum density) and $n=1.9$ (Cyan with minimum density). The specific values of constant parameters are mentioned in Table-\ref{tab1}.\{$n=1 (\textcolor{black}{\bigstar})$, $n=1.2(\textcolor{purple}{\bigstar})$, $n=1.4\;(\textcolor{magenta}{\bigstar})$, $n=1.5(\textcolor{red}{\bigstar})$, $n=1.6(\textcolor{orange}{\bigstar})$, $n=1.7(\textcolor{violet}{\bigstar})$, $n=1.8\;(\textcolor{blue}{\bigstar})$, and $n=1.9(\textcolor{cyan}{\bigstar})$.\}}
\end{figure}\label{fig4}

\begin{figure}
\includegraphics[width=70mm,height=50mm]{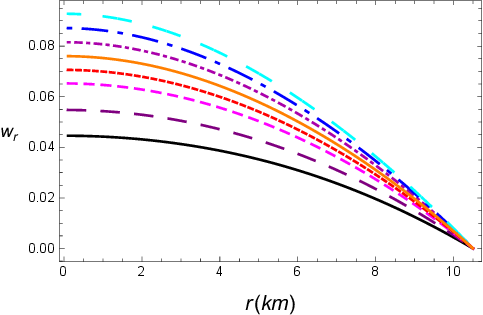}\\\includegraphics[width=70mm,height=55mm]{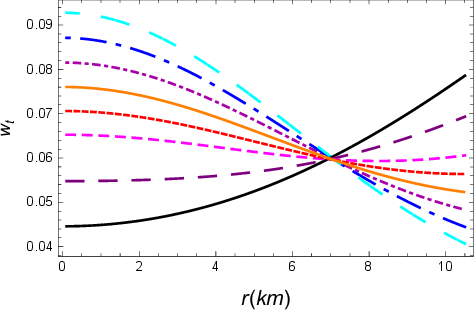}\\\includegraphics[width=70mm,height=55mm]{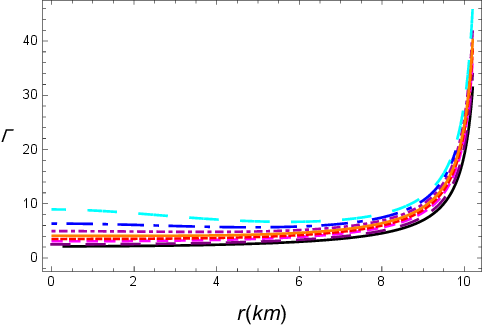}
\caption{Behaviors of the equation of state components (radial and tangential), and adiabatic index versus radial coordinate $r$.Here $n=1.0$ (Black with maximum density) and $n=1.9$ (Cyan with minimum density). The specific values of constant parameters are mentioned in Table-\ref{tab1}.\{$n=1 (\textcolor{black}{\bigstar})$, $n=1.2(\textcolor{purple}{\bigstar})$, $n=1.4\;(\textcolor{magenta}{\bigstar})$, $n=1.5(\textcolor{red}{\bigstar})$, $n=1.6(\textcolor{orange}{\bigstar})$, $n=1.7(\textcolor{violet}{\bigstar})$, $n=1.8\;(\textcolor{blue}{\bigstar})$, and $n=1.9(\textcolor{cyan}{\bigstar})$.\}}
\end{figure}\label{fig5}

\subsection{Behavior of Metric Potentials, energy density, pressure profiles, and gradients}

Here we analyze the graphical behavior of the metric potentials $\rho$, $p_r$, and $p_t$ using the mass and radii of the considered compact star, namely $PSR J0437-4715$. The behavior of the metric components is depicted in Fig. \ref{9}. It is evident through these metric components that the presented model satisfies the condition $e^{\lambda(r=0)}=1$ and $e^{\nu(r=0)}\neq0$, making it physically acceptable. The graphical illustrations demonstrate that both metric components exhibit a monotonically increasing and regular behavior, particularly at the center, in the interior region of the star.

The smooth and regular behavior of the density is given in Fig. \ref{9}, showing that the matter density reaches its maximum at the center of the compact star ($r=0$) and decreases gradually as the radius increases towards the outer surface. Fig. \ref{1} illustrates the behavior of the radial and tangential pressures ($p_r$ and $p_t$) which indicates that they attain maximum values at the center and decrease gradually near the boundary of the compact star. It is also observed that the density and pressures vanish at some finite radius.

Furthermore, the gradients of these quantities are illustrated in Fig. \ref{2}, where all the respective quantities exhibit a negative decreasing behavior. Thus the constraints $\frac{d \rho}{dr}<0$, $\frac{d p_r}{dr}<0$, and $\frac{d p_t}{dr}<0$ are satisfied.

\subsection{ Anisotropic Nature of Stellar Structure}

For a spherically symmetric metric, the anisotropy parameter $\Delta$ is determined by the relation $\Delta=(P_t-P_r)$. The graphical representation of this parameter is shown in Fig. \ref{2} which reveals an increasing trend. The values of $\Delta$ appear positive for all three models of the star, indicating the presence of repulsive forces that balance the gravitational forces within the massive distribution. It is noteworthy that the anisotropy vanishes at the center of the stellar structure.

\subsection{Energy Conditions}
The validity of energy bounds has been the fundamental criterion for establishing a physically realistic and viable star model. In this subsection, we examine the behavior of these constraints graphically for the stellar structure. In general, these conditions are expressed in terms of inequalities involving matter density and pressures, known as the null energy condition (NEC), weak energy condition (WEC), strong energy condition (SEC), and dominant energy condition (DEC). Mathematically, they are defined as follows:
\begin{eqnarray}
\begin{aligned}
 NEC: \rho + p_r \geq 0, \quad \rho + p_t \geq 0,&&\\
 WEC: \rho \geq 0, \quad \rho + p_r \geq 0, \quad \rho + p_t \geq 0,&&\\
 SEC: \rho + p_r \geq 0, \quad \rho + p_t \geq 0, \quad \rho + p_r + 2p_t \geq 0,&&\\
 DEC: \rho > |p_r|, \quad \rho > |p_t|.&&
\end{aligned}
\end{eqnarray}
The graphical representation of the energy conditions is provided in Fig. \ref{3}. It can be observed that all the energy bounds are satisfied for the considered compact star models. This confirms that the presented stellar structure is physically valid.

\subsection{Equation of state components}

In addition, we investigate the graphical behavior of radial and tangential equations of state (EoS) parameters, denoted by $\omega_r$ and $\omega_t$, respectively (Fig. \ref{4}). The graphical analysis clearly demonstrates that both EoS parameters,  $\omega_r$ and $\omega_t$, satisfy the constraints $\omega_t>0$ and $0<\omega_r<1$. This behavior of the EoS parameters confirms the realism of the stellar object model.

\subsection{Adiabatic Index}
The adiabatic index is a crucial parameter that characterizes the stability of a stellar structure. In the case of a spherically symmetric object, the study of the adiabatic index becomes particularly important as it reflects its solidity. It has been suggested that the adiabatic index should satisfy the inequality $\Gamma>\frac{4}{3}$ within the interior of a compact star to ensure stability \cite{60}.

Generally, the adiabatic index is defined as
\begin{equation}
\Gamma=\frac{\rho + p_r}{p_r}\frac{d p_r}{d \rho}.
\end{equation}
In our current analysis, we study the graphical behavior of the adiabatic index, as shown in Fig. \ref{4}. It is found that the adiabatic index starts from $0$ at the center of the star and then increases monotonically. This behavior satisfies the required stability criterion ($\Gamma>\frac{4}{3}$) at the surface of the star.

\subsection{Zeldovich's Condition}
Zeldovich's condition establishes a criterion for determining the stability of a stellar object configuration based on the central values of energy density $\rho_c$ and radial pressure $p_{rc}$ at $r=0$. According to this condition, a stellar structure is considered stable if the following constraint on the ratio holds.
\begin{equation}
\frac{p_{rc}}{\rho_c}<1.
\end{equation}
In this analysis, we examine this condition mathematically whose results are presented in Table-\ref{tab1}. It is observed that Zeldovich's condition holds, confirming the stability of the constructed model.
\subsection{Stability and Equilibrium}
This subsection is devoted to discussing the graphical representation of casuality and equilibrium of the stellar model using TOV equations as well as the velocities of sound in the interior region of a compact star.

\subsubsection{Equilibrium Under Various Forces}
In the literature, the Tolman-Oppenheimer-Volkoff (TOV) equation is recognized as an equilibrium condition for a strange star, accounting for gravitational, hydrostatic, and anisotropic forces \cite{53, 54}. The general form of the TOV equation is given by
\begin{eqnarray}
\frac{d p_r}{dr}+\frac{\nu^{'}(\rho+p_r)}{2}-\frac{2(p_t-p_r)}{r}=0.
\end{eqnarray}
Alternatively, it can be expressed as
\begin{eqnarray*}
F_g+F_h+F_a=0; \quad F_g=-\frac{\nu^{'}(\rho+p_r)}{2}, \quad F_h=-\frac{d p_r}{dr},\\ \quad F_a=\frac{2(p_t-p_r)}{r}.
\end{eqnarray*}
Here $F_g$, $F_h$, and $F_a$ represent the gravitational, hydrostatic, and anisotropic forces, respectively. Putting the energy density, radial pressure, and tangential pressure into the TOV equation, we draw these forces graphically as shown in Fig.\ref{2}. We examine that these forces individually exhibit small positive or negative values, but collectively, they satisfy the TOV equation. Thus we conclude that the presented stellar structure is in a state of equilibrium under the influence of these forces.
\begin{figure}
\includegraphics[width=70mm,height=50mm]{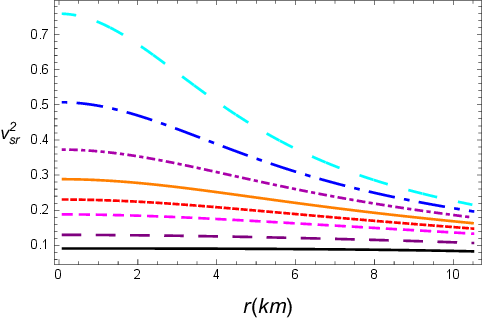}\\\space\includegraphics[width=70mm,height=50mm]{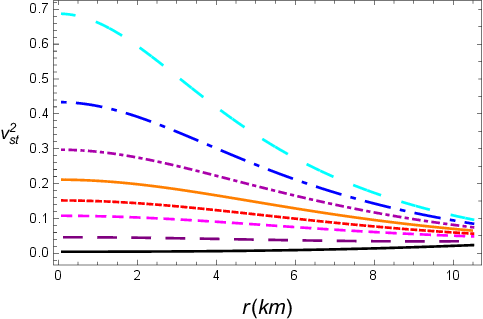}\\\includegraphics[width=73mm,height=55mm]{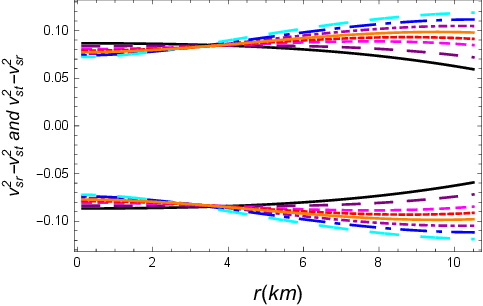}
\caption{The speeds of sound and Abreu condition versus radial coordinate $r$. Here $n=1.0$ (Black with maximum density) and $n=1.9$ (Cyan with minimum density). The specific values of constant parameters are mentioned in Table-\ref{tab1}.\{$n=1 (\textcolor{black}{\bigstar})$, $n=1.2(\textcolor{purple}{\bigstar})$, $n=1.4\;(\textcolor{magenta}{\bigstar})$, $n=1.5(\textcolor{red}{\bigstar})$, $n=1.6(\textcolor{orange}{\bigstar})$, $n=1.7(\textcolor{violet}{\bigstar})$, $n=1.8\;(\textcolor{blue}{\bigstar})$, and $n=1.9(\textcolor{cyan}{\bigstar})$.\}}
\end{figure}\label{fig6}
\begin{figure}
\includegraphics[width=70mm,height=50mm]{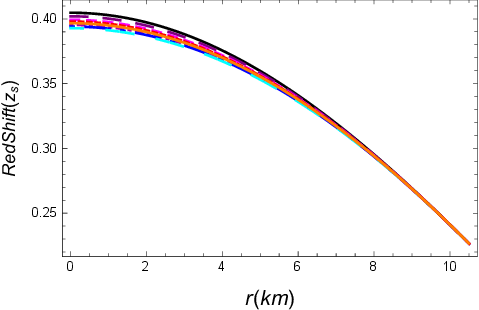}
\caption{The gravitational redshift versus radial coordinate $r$. Here $n=1.0$ (Black with maximum density) and $n=1.9$ (Cyan with minimum density). The specific values of constant parameters are mentioned in Table-\ref{tab1}.\{$n=1 (\textcolor{black}{\bigstar})$, $n=1.2(\textcolor{purple}{\bigstar})$, $n=1.4\;(\textcolor{magenta}{\bigstar})$, $n=1.5(\textcolor{red}{\bigstar})$, $n=1.6(\textcolor{orange}{\bigstar})$, $n=1.7(\textcolor{violet}{\bigstar})$, $n=1.8\;(\textcolor{blue}{\bigstar})$, and $n=1.9(\textcolor{cyan}{\bigstar})$.\}}
\end{figure}\label{fig7}

\subsubsection{Causality and Stability Condition}

In this study, we explore the validity of the causality conditions for the constructed stellar model. The causality conditions impose restrictions on the radial and tangential velocities denoted by $v^2_r$ and $v^2_t$, respectively. These constraints are defined as $0<\mid v^2_i\mid<1$, where $i=r,t$. The radial and transverse velocities are given by
\begin{eqnarray}
\begin{aligned}
v^2_r=\frac{d p_r}{d\rho}, \quad v^2_t=\frac{d p_t}{d\rho}.
\end{aligned}
\end{eqnarray}

The graphical behavior of these velocities is depicted in Fig. \ref{5}, indicating the validity of causality conditions. Additionally, another stability condition based on these velocities has been proposed by Abreu \cite{55}, defined as $-1\leq v^{2}_{t} - v^{2}_{r}\leq1$. This condition is considered an interesting feature of the compact star models. In our case, we also examine this condition graphically as shown in Fig. \ref{5}. The graph clearly demonstrates that our model is consistent with this stability condition as well. In conclusion, we can state that our model is stable for all positive values of $n\leq2.0$.

\subsection{Redshift, Compactness, and Mass Function}

The core of a compact star is characterized by the compactness factor, which describes the relativistic nature of the star. The compactness is represented by $\mu(r)$ and is defined as the ratio of the mass function $m(r)$ to the radius of the star $r$. The mass function $m(r)$ is defined as
\begin{eqnarray}
m(r)=\frac{1}{2}\int \rho(r)r^2dr.
\end{eqnarray}
Moreover, the compactness factor $u(r)$ can be defined using the mass function 
\begin{eqnarray}\nonumber
u&=&\frac{M}{R}, ~~\text{where}~~ M=\frac{m(R)}{\alpha}\\
\end{eqnarray}
while surface redshift is determined by the formula, 
\begin{eqnarray}\nonumber
z_s&=&e^{-\frac{\nu(R)}{2}}-1 =(1-2u)^{-1/2}-1.
\end{eqnarray}
This behavior of gravitational redhsift is shown in Fig. \ref{6}.

\begin{figure*}
\centering \includegraphics[width=9cm,height=6.5cm]{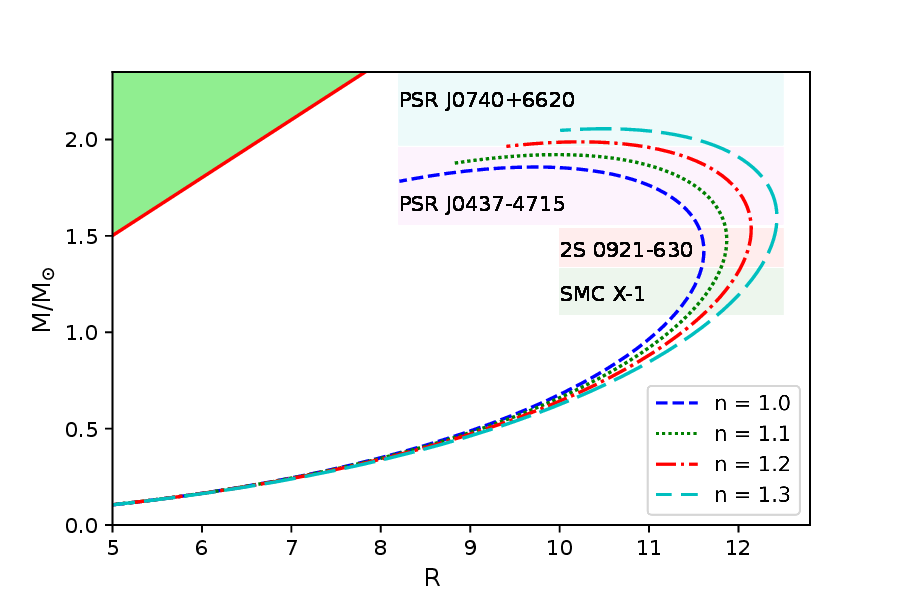}\includegraphics[width=9cm,height=6.5cm]{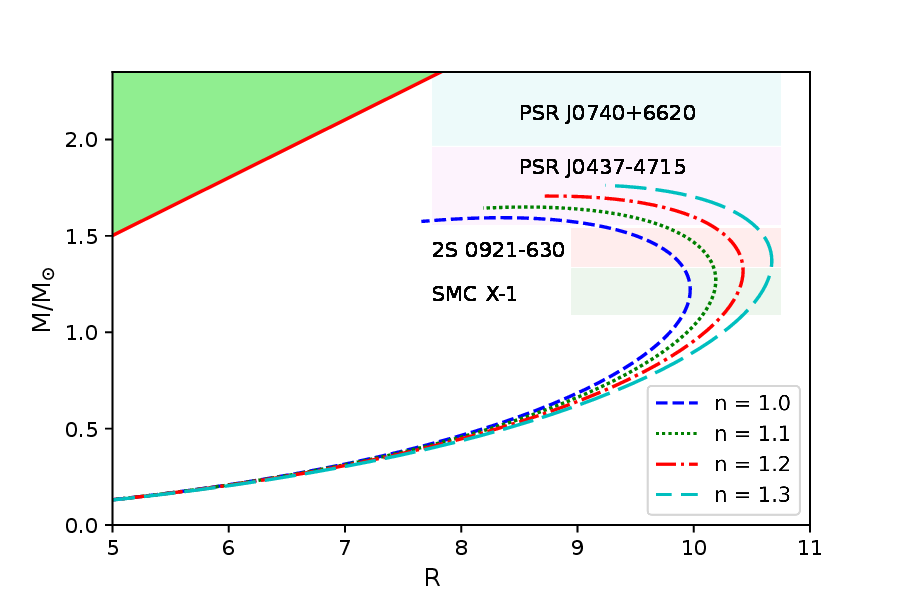}
\caption{Graphical representation of Mass-Radius curves for the surface densities $4.55\times 10^{14} gm/cm^{3}$ (left panel) and $5.35\times 10^{14} gm/cm^{3}$ (right panel) by taking $\alpha = -0.65,~\beta= 2.036\times10^{-35},~a = 0.3,~b = 0.0015,~B=0.03876$ and the four sets of parameters \{n, A, c\}, i.e., \{1.0, -13.930, 0.00063\}, \{1.1, -15.540, 0.00045\}, \{1.2, -17.790, 0.00032\}, \{1.3, -20.938, 0.00023\}. } \label{MRn}
\end{figure*}

\begin{figure*}
\centering \includegraphics[width=9cm,height=6.5cm]{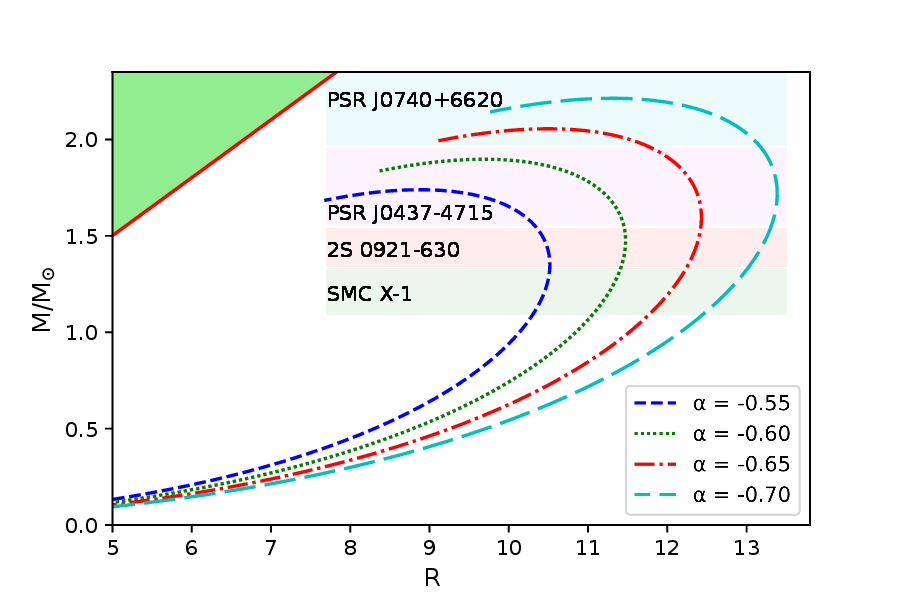}\includegraphics[width=9cm,height=6.5cm]{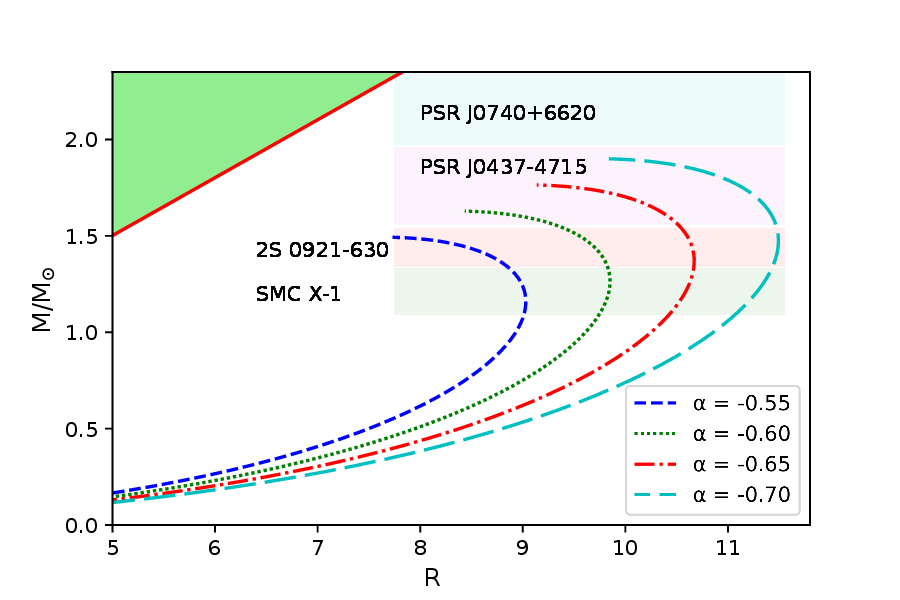}
\caption{Graphical representation of Mass-Radius curves for the surface densities $4.55\times 10^{14} gm/cm^{3}$ (left panel) and $5.35\times 10^{14} gm/cm^{3}$ (right panel) by taking $\beta= 2.036\times10^{-35}, ~a = 0.3,~b = 0.0015,~B=0.03876$ and the set of parameters \{n, A, c\}, i.e., \{1.3, -20.938, 0.00023\}.} \label{MRalpha}
\end{figure*}

\begin{table*}
\centering \caption{Predicted radii of different stellar candidates from the model for $\alpha=-0.65$}
\scalebox{0.92}
 {\begin{tabular}{cccccccccc}
\hline
\multirow{2}{*}{Stellar objects} & \multirow{2}{*}{${\frac{M}{M_\odot}}$}  & \multicolumn{4}{c}{{Predicted radii (km) for $\rho_s\simeq 4.55\times 10^{14} gm/cm^{3}$}} & \multicolumn{4}{c}{{Predicted radii (km) for $\rho_s\simeq 5.35\times 10^{14} gm/cm^{3}$}} \\
\cline{3-10} 
&& $n=1.0$ &  $n=1.1$ &   $n=1.2$  &  $n=1.3$ & $n=1.0$ & $n=1.1$ & $n=1.2$ & $n=1.3$ \\ \hline
 PSR J0740+6620 \cite{Cromartie2020} & $2.14^{+0.20}_{-0.18}$  
&  -                        &  -                        & 10.85                     &  11.72 
&   -                       &      -                    & -                         & - \\ \hline
PSR J0437-4715 \cite{Verbiest2008}  & $1.76\pm0.20$   
& 11.01-11.54               & 11.51-11.85               & $11.94_{+0.24}^{-0.96}$   &   $12.33_{+0.10}^{-0.55}$  
& 9.12                      & 9.69                      & 10.13                     & 9.33-10.51\\ \hline
2S 0921-630 \cite{Steeghs2007}  &  $1.44\pm0.10$    
&  $11.69_{-0.02}^{-0.05}$  & $11.86_{-0.05}^{-0.01}$   &  $12.11_{-0.08}^{+0.03}$  & $12.36^{+0.06}_{-0.11}$  
&   $9.72^{-0.44}_{+0.18}$  &  $10.05^{-0.27}_{+0.12}$  &  $10.36^{-0.18}_{+0.06}$  & $10.65^{-0.10}_{+0.02}$\\ \hline
SMC X-1 \cite{Falanga2015}  & $1.21\pm0.12$  
& $11.47^{+0.11}_{-0.19}$   & $11.66^{+0.14}_{-0.22}$   &  $11.85^{+0.17}_{-0.25}$  & $12.04^{+0.20}_{-0.27}$ 
& $9.97^{-0.05}_{-0.06}$    & $10.18^{-0.01}_{-0.10}$   & $10.38^{+0.04}_{-0.13}$   & $10.59^{+0.08}_{-0.16}$\\ \hline
  \label{table2}
\end{tabular} }
\end{table*}
\begin{table*}
\centering \caption{Predicted radii of different stellar candidates from the model for $n = 1.3$}
\scalebox{0.92}
 {\begin{tabular}{cccccccccc}
\hline
\multirow{2}{*}{Stellar objects} & \multirow{2}{*}{${\frac{M}{M_\odot}}$}  & \multicolumn{4}{c}{{Predicted radii (km) for $\rho_s\simeq 4.55\times 10^{14} gm/cm^{3}$}} & \multicolumn{4}{c}{{Predicted radii (km) for $\rho_s\simeq 5.35\times 10^{14} gm/cm^{3}$}} \\
\cline{3-10} 
&& $\alpha=-0.55$ &  $\alpha=-0.60$ &   $\alpha=-0.65$  &  $\alpha=-0.70$ & $\alpha=-0.55$ & $\alpha=-0.60$ & $\alpha=-0.65$ & $\alpha=-0.70$ \\ \hline
 PSR J0740+6620 \cite{Cromartie2020} & $2.14^{+0.20}_{-0.18}$  
&  -                        &  -                        & 11.73                     &  12.51-13.15
&   -                       &      -               & -                      & -\\ \hline
PSR J0437-4715 \cite{Verbiest2008}  & $1.76\pm0.20$   
& 10.31              & 11.08-11.44               & $12.33_{+0.11}^{-0.55}$   &   $13.38_{-0.06}^{+0.21}$  
&   -                       &      9.30                   & 9.33-10.51                       & 11.11-11.47\\ \hline
2S 0921-630 \cite{Steeghs2007}  &  $1.44\pm0.10$    
&  $10.48_{+0.03}^{-0.13}$  & $11.47_{-0.05}^{-0.02}$   &  $12.36_{-0.11}^{+0.06}$  & $13.19^{+0.11}_{-0.16}$  
&   8.46-8.85  &  $9.70^{-0.30}_{+0.12}$  &  $10.65^{-0.11}_{+0.02}$  & $11.49^{-0.01}_{-0.05}$\\ \hline
SMC X-1 \cite{Falanga2015}  & $1.21\pm0.12$  
& $10.45^{+0.06}_{-0.15}$   & $11.27^{+0.14}_{-0.21}$   &  $12.04^{+0.20}_{-0.27}$  & $12.76^{+0.25}_{-0.31}$ 
& $9.02^{-0.15}_{-0.01}$    & $9.84^{-0.01}_{-0.10}$   & $10.59^{+0.07}_{-0.17}$   & $11.29^{+0.14}_{-0.22}$\\ \hline
  \label{table3}
\end{tabular} }
\end{table*}

\section{Maximum Mass, Radius and different features of Mass-Radius curves in $f(Q)$ gravity}

 In this section, we study the Mass-Radius relationship that is considered to be an important aspect of the present model for its physical acceptance. The behavior of the $M-R$ curves with respect to different values of $n$ and $\alpha$ is plotted in Figs. \ref{MRn} and \ref{MRalpha} corresponding to two different densities, i.e., $4.55\times 10^{14} gm/cm^{3}$ and $5.35\times 10^{14} gm/cm^{3}$. We display four stellar candidates of the known observed masses in Figs. \ref{MRn} and \ref{MRalpha} and also estimate radii of the stars from the $M-R$ curves. The predicted radii subject to the chosen values of parameters $\{\beta, ~a, ~b,~c\}$  are enlisted in Tables \ref{table2} and \ref{table3}. The variation of maximum mass and the corresponding radius is increasing in nature with respect to increasing values of $n$ and decreasing values of $\alpha$, respectively. Thus the maximum observed masses for different stars can be justified with increasing degree of the power $n$ present in metric potentials, i.e., embedding class-I solutions in $f(Q)$ gravity with $\alpha$ having the direct influence on the maximum mass. 

\begin{table*} 
\centering \caption{Maximum mass and Radius from M-R curves for different values of $n$}
\begin{tabular}{cccccccccc}
\hline
 \multirow{2}{*}{$n$}  & \multicolumn{5}{c}{{For $\rho_s\simeq 4.55\times 10^{14} gm/cm^{3}$}} & \multicolumn{4}{c}{{For $\rho_s\simeq 5.35\times 10^{14} gm/cm^{3}$}} \\
\cline{2-10} 
&& $M_{max}/M_{\odot}$ &  $M_{max}$ (km) &   $R$ (km)  &  $M/R$ & $M_{max}/M_{\odot}$ & $M_{max}$ (km) & $R$ (km) & $M/R$ \\ \hline
1.0  &  
&  1.86                      &  2.75                      & 9.73                  &  0.283
&   1.59                       &     2.35                   & 8.35                      & 0.281 \\ \hline
1.1 & 
& 1.92               & 2.83             & 9.97   &   0.284 
& 1.65                      & 2.44                    & 8.56                    & 0.285 \\ \hline
1.2  &    
&  1.99  & 2.94  &  10.23  & 0.287 
&   1.71  &  2.52  &  8.78  & 0.287\\ \hline
1.3 &  
& 2.06   & 3.04   &  10.50  & 0.290
& 1.76    & 2.60   & 9.24   & 0.281\\ \hline
  \label{table4}
\end{tabular} 
\end{table*}
\begin{table*}
\centering \caption{Maximum mass and Radius from M-R curves for different values of $\alpha$}
\begin{tabular}{cccccccccc}
\hline
 \multirow{2}{*}{$\alpha$}  & \multicolumn{5}{c}{{For $\rho_s\simeq 4.55\times 10^{14} gm/cm^{3}$}} & \multicolumn{4}{c}{{For $\rho_s\simeq 5.35\times 10^{14} gm/cm^{3}$}} \\
\cline{2-10} 
&& $M_{max}/M_{\odot}$ &  $M_{max}$ (km) &   $R$ (km)  &  $M/R$ & $M_{max}/M_{\odot}$ & $M_{max}$ (km) & $R$ (km) & $M/R$ \\ \hline
-0.55  &  
&  1.74                    &  2.57                      & 8.89                 &  0.289
&   1.49                       &     2.20                   & 7.73                     & 0.285 \\ \hline
-0.60 & 
& 1.90              & 2.80             & 9.69   &   0.281
& 1.63                     & 2.41                   & 8.43                   & 0.286 \\ \hline
-0.65  &    
&  2.06  & 3.04  &  10.50  & 0.290 
&   1.76  &  2.60 &  9.24  & 0.281\\ \hline
-0.70 &  
& 2.21   & 3.26   &  11.31  & 0.288
& 1.90   & 2.80   & 9.84   & 0.285\\ \hline
  \label{table5}
\end{tabular} 
\end{table*}

We analyze the M-R relationship in the context of $f(Q)$ gravity by extracting points of maximum mass and radius from each curve of the Figs. \ref{MRn} and \ref{MRalpha}. These numerical data points are tabulated in Tables \ref{table4} and \ref{table5}. Eventually, we find that maximum mass and radius are reduced significantly for increasing changes in the surface densities keeping all other parameters fixed. However, an increase in surface densities with other fixed parameters causes a nominal change in the compactification factor, i.e., $M/R$ within the Buchdahl limit.  It is worth mentioning that stellar objects of high compactness can be obtained in $f(Q)$ gravity for different choices of geometric parameters $\{a, ~b,~c\}$ of embedding class I spacetime.

In the present analysis, PSR J0437-4715 \cite{Demorest2010} is considered as a reference to study the physical validity of the compact stars using class I solutions in $f(Q)$ gravity. From the previous analysis of M-R curves, we see that PSR J0437-4715 is the stellar candidate best described by the M-R curve specific to $n = 1.2$ and $\alpha=-0.65$ with surface density $4.55\times 10^{14} gm/cm^{3}$. Further, to ensure the practical validity of our model, we consider observational data for four stellar candidates such as  PSR J0740+6620 \cite{Cromartie2020}, PSR J0437-4715 \cite{Verbiest2008}, 2S 0921-630 \cite{Steeghs2007} and SMC X-1 \cite{Falanga2015}. In a recent study \cite{Miller2021}, the radius of PSR J0740+6620 is determined to be $13.7^{+2.6}_{-1.5}$ with 68\% credibility using observational X-ray data from NICER and XMM-Newton. This result seems well consistent with the predicted radius of PSR J0740+6620, i.e.,  12.51-13.15 km for $\alpha=-0.70,~n = 1.3$ as shown in Table \ref{table2}. The upper limit of the observed radius of PSR J0740+6620 can be found from the present model for smaller values of $\alpha$. Miller et al. \cite{Miller2021} measured range of radius for 1.4 $M_{\odot}$ neutron as $12.45\pm 0.65$ km  considering density equal to $1.5 -3 \times \rho_{sat}$, where $\rho_{sat} = 2.7-2.8 \times 10^{14} gm/cm^{3}$ is the nuclear saturation density. This is in good accordance with the present model as Table \ref{table3} suggests that 2S 0921-630 (1.44 $M_{\odot}$) have the range of radius 11.47-13.19 km for $-0.70<\alpha<-0.60$, $n=1.3$ and  $\rho_s \approx 1.65 \times \rho_{sat}$. In another study \cite{Denis2019}, the measured radius $13.6^{+0.9}_{-0.8}$ km of PSR J0437-4715 using ultraviolet and X-ray thermal emission data is in close agreement with $13.38_{-0.06}^{+0.21}$ km for $\alpha=-0.70,~n=1.3$ (see Table \ref{table2}). Finally, the detailed analysis presented above proposes that the observational limits of radii of different compact stars are likely to follow the $M-R$ curve of the present model for $\alpha=-0.70,~n=1.3$ and $\rho_s \approx 1.65 \times \rho_{sat}$. Hence, this conclusion is tempting to predict the radius of SMC X-1 to be $12.76^{+0.25}_{-0.31}$ km (see Table \ref{table2}) which may be in concurrence with the observed radius range of SMC X-1 if determined experimentally in future.

\section{Conclusion}\label{sec6}
In this work, we have investigated the spherically symmetric compact star model with an anisotropic fluid. We considered the $f(Q)$ generic function of the form $f(Q)=\alpha Q+\beta$, where $\alpha$ is a small arbitrary constant and $\beta$ represents the cosmological constant. The Karmarkar condition has been applied to obtain the differential equation relating the metric potentials of the spacetime. Solving this differential equation, we have derived an expression of the metric component $e^\nu$ in terms of $e^\lambda$. The comparison of the interior spherical metric with the exterior Schwarzschild metric led to constraints on the arbitrary constants in this configuration. The obtained results were presented graphically in Figs. \ref{9}-\ref{6}. The key findings can be summarized as follows:
\begin{itemize}
\item The metric potentials exhibit regular and monotonically decreasing behavior with respect to the radial coordinate.
\item The matter density and pressures show a regular decreasing behavior, reaching maximum values at the center of the star.
\item The radial and tangential equations of state (EoS) parameters satisfy the inequalities $0<\mid v^2_r,~v^2_t \mid<1$. They also satisfy the Abreu constraint: $-1\leq v^{2}_{t} - v^{2}_{r}\leq1$.
\item The anisotropy parameter $\Delta$ diminishes at the center of the star and is greater than zero everywhere inside the star, indicating the presence of repulsive forces.
\item The energy conditions have been satisfied based on the graphical analysis.
\item The stability of the constructed model is ensured by examining the Tolman-Oppenheimer-Volkoff (TOV) equation and causality conditions graphically.
Zeldovich's condition has been verified mathematically and found to be consistent with the present model.

\item  The redshift satisfies the Bowers and Liang condition $Z_s<5$, and the compactness parameter validates the Buchdahl condition $u(r)<4/9$. According to new study \cite{alho2022compactness}, the compactness limit can be reached up to  $u(r)\approx 0.462$, which is beyond the standard proposed maximum allowable compactness limit for static elastic object satisfying causality conditions. The compactness for our present models is less than the Buchdahl limit, and hence our models automatically fulfill the newly proposed limit. 

\item The variation in maximum mass and the corresponding radius increases as $n$ values increase and $\alpha$ values decrease, respectively. Consequently, the justification for the maximum observed masses in different stars can be linked to the increasing power of $n$ in metric potentials. This occurs specifically in the context of embedding class-I solutions in $f(Q)$ gravity, with $\alpha$ directly influencing the maximum mass. We observe a significant reduction in both maximum mass and radius with increasing changes in surface densities while keeping all other parameters constant. However, when surface densities increase with fixed parameters, there is only a nominal change in the compactification factor, i.e., $M/R$, within the Buchdahl limit. It is noteworthy that high compactness stellar objects can be achieved in $f(Q)$ gravity for different values of geometric parameters $A, B, c$ in the embedding class-I spacetime.
\end{itemize}

Additionally, we have investigated the behavior of various quantities such as metric potentials, matter density, radial and tangential pressures, energy density and pressure gradients with respect to the radial coordinate, radial and tangential equations of state parameters, anisotropy parameter, energy conditions, Zeldovich's condition, stability, causality condition, adiabatic index, redshift, compactness, and mass function for the astronomical objects PSR J0437-4715. The results are summarized in Table-\ref{tab1} which support the physical viability and mathematical consistency of our spherically symmetric model.

\section*{Acknowledgments}
The paper was funded by the National Natural Science Foundation of China 11975145.

\end{document}